\documentclass[]{tCPH2e}
\usepackage{natbib}
\newcommand{\deriv}{\mathrm{d}}

\newcommand{\simless}{\mathbin{\lower 3pt\hbox
   {$\rlap{\raise 5pt\hbox{$\char'074$}}\mathchar"7218$}}}
\newcommand{\simgreat}{\mathbin{\lower 3pt\hbox
   {$\rlap{\raise 5pt\hbox{$\char'076$}}\mathchar"7218$}}}

\begin{document}
\doi{10.1080/0010751YYxxxxxxxx}
 \issn{1366-5812}
\issnp{0010-7514}

\jvol{00} \jnum{00} \jyear{2014} \jmonth{January}

\markboth{Taylor \& Francis and I.T. Consultant}{Contemporary Physics}


\title{Is the Universe Simpler Than $\Lambda$CDM?}

\author{Matthew G. Walker$^{a,b}$$^{\ast}$\thanks{$^\ast$Corresponding author. Email: mgwalker@cmu.edu
\vspace{6pt}} and Abraham Loeb$^{b}$\\\vspace{6pt}  $^{a}${\em{McWilliams Center for Cosmology, Carnegie Mellon University, 5000 Forbes Ave., Pittsburgh, PA 15213}};
$^{b}${\em{Institute for Theory and Computation, Harvard University, 60 Garden St., Cambridge, MA 02138}
}\\\vspace{6pt}\received{v3.0 released January 2010} }

\maketitle

\begin{abstract}
In the standard cosmological model, the Universe consists mainly of two invisible substances: vacuum energy with constant mass-density $\rho_{\rm v}=\Lambda/(8\pi G)$ (where $\Lambda$ is a `cosmological constant' originally proposed by Einstein and $G$ is Newton's gravitational constant) and cold dark matter (CDM) with mass density that is currently $\rho_{\rm DM,0}\sim 0.3\rho_{\rm v}$.  This `$\Lambda$CDM' model has the virtue of simplicity, enabling straightforward calculation of the formation and evolution of cosmic structure against the backdrop of cosmic expansion.  Here we review apparent discrepancies with observations on small galactic scales, which $\Lambda$CDM must attribute to complexity in the baryon physics of galaxy formation.  Yet galaxies exhibit structural scaling relations that evoke simplicity, presenting a clear challenge for formation models.  In particular, tracers of gravitational potentials dominated by dark matter show a correlation between orbital size, $R$, and velocity, $V$, that can be expressed most simply as a characteristic acceleration, $a_{\rm DM}\sim 1$ km$^2$s$^{-2}$pc$^{-1}\approx3\times 10^{-9}$ cm s$^{-2}\approx 0.2c\sqrt{G\rho_{\rm v}}$, perhaps motivating efforts to find a link between localized and global manifestations of the Universe's dark components.

\bigskip
\bigskip
\end{abstract}

\section{Introduction}

Gravity regulates the Universe's expansion and the growth of its structure.  Observations of both phenomena reveal accelerations that cannot be attributed to classical gravitational fields sourced by known particles.  Rather than decreasing as galaxies and galaxy clusters gravitationally attract each other, the rate at which the Universe expands is \textit{increasing}.  Inside those galaxies and galaxy clusters, orbiting bodies (stars within galaxies, individual galaxies within galaxy clusters) reach speeds in excess of escape velocities inferred from the amount of visible material, yet remain gravitationally bound.  

\smallskip Both results imply new physics, requiring revision of either the Universe's composition or its laws.  The acceleration of cosmic expansion \citep{riess98,perlmutter99} requires either a new substance that induces gravitational repulsion \citep{weinberg13} or a modification of general relativity that becomes apparent only on large scales \citep[][]{carroll04}.  Galactic dynamics---or, more generally, the large ratios of dark to luminous mass inferred for gravitationally bound structures on galactic and larger scales \citep[][]{zwicky37}---require either a new substance that interacts almost exclusively via gravity or a modification of general relativity that becomes apparent only in regions of weak acceleration \citep[][]{milgrom83}.  

\smallskip
`Dark energy' and `dark matter' refer generically to the substance-based interpretations.  Both categories admit various candidates that can be distinguished in terms of their complexity.  The current cosmological paradigm is built on the hypothesis that both substances take extremely simple forms, such that their influence on cosmic evolution can be calculated given the values of a few quantities that specify initial conditions \citep{spergel03,hinshaw13,planck13}.  In this model, dark energy reduces to a constant mass-density of vacuum, $\rho_{\rm v}=\Lambda /(8\pi G)$, effectively exerting negative pressure, $P=-\rho_{\rm v}$, that makes cosmic acceleration positive (i.e., $\Lambda$ plays the role of the `cosmological constant' that Einstein originally introduced in order to prevent the gravitational collapse of a universe he presumed to be static).  Dark matter reduces to a fluid made of `cold', `collisionless' particles that form with negligible velocity dispersion and avoid non-gravitational interactions, letting CDM particles clump together gravitationally on small---i.e., subgalactic---scales.  

\smallskip
It is this ability that distinguishes CDM\footnote{Throughout this article, by `CDM' we mean any dark matter particle candidate for which primordial velocity dispersions and non-gravitational interactions have negligible impact on galaxy formation.} astrophysically from alternatives like `warm', `hot', or `self-interacting' dark matter.  In these alternative scenarios, significant primordial velocity dispersions and/or non-gravitational scattering mechanisms suppress clustering below a characteristic scale that is sufficiently large to affect the properties of observed galaxies.  It may therefore be possible to test the CDM hypothesis by using observations of galactic structure to look for evidence that dark matter has a minimum clustering scale.  Here we review efforts to find such a scale and discuss possible implications for the $\Lambda$CDM model.

\section{Toward Small Scales}
\label{sec:scales}

\smallskip
The attribution of cosmic acceleration to small but non-zero vacuum energy, $\rho_{\rm v}\sim 7\times 10^{-30}~{\rm g~cm^{-3}}$, remains straightforwardly consistent with all available data, including 1) luminosities measured to redshifts\footnote{Here `redshift' refers to the Doppler shift of light from distant objects toward redder wavelengths as the Universe expands.  The expansion is generally described in terms of the growth of a dimensionless scale factor, $a$, which is related to redshift, $z$, by $a^{-1}=1+z$.} $z\simless 1.5$ for standard `candles' like Type-Ia supernovae \citep[][]{riess07}, 2) angular-diameters measured to redshifts $z\simless 2$ for standard `rulers' like the clustering of baryons in response to standing sound waves in the early universe \citep[][]{mehta12}, 3) the global geometry and rate of structure growth as inferred from weak gravitational lensing \citep[][]{mandelbaum13} and/or abundances of galaxy clusters detected in optical, X-ray and microwave surveys \citep[][]{vikhlinin09}; for a detailed review, see reference \cite{weinberg13}.  

\smallskip 
The situation is different for CDM.  While many of the same cosmological observations that signal dark energy similarly imply non-baryonic dark matter, astrophysical tests that might distinguish CDM from other viable particle candidates tend to require examination of the smaller scales that characterise gravitationally-collapsed, galactic structure.  CDM's simplicity then has practical value, as the absence of a cosmologically relevant scale for CDM clustering enables fast numerical simulations of cosmic structure formation \citep[][]{frenk12}, providing an invaluable tool for visualising and quantifying predictions of the CDM hypothesis.  Over the past 25 years, $\Lambda$CDM simulations have developed rapidly as 1) initial conditions have been tuned to match cosmological parameters measured with ever-increasing precision and 2) numerical resolution has improved in lock step with computational speed.

\smallskip
Results now include robust predictions about cosmic structure formation, at least in the case that dissipative non-gravitational forces are negligible.  CDM clustering can begin at or below \textit{solar-system} scales, with gravitational collapse of `microhalos' of $\simless 1$ Earth mass and size smaller than the distance between Earth and Sun \citep{green04}.  Structure formation then proceeds `hierarchically', with larger halos assembled from mergers and accretion of smaller ones, such that the halo of a Milky-Way-like galaxy hosts trillions of subhalos (and sub-subhalos, etc. \cite{diemand05}).  On the largest scales, halos populate thin filaments that join at dense nodes, threading large voids to form a cosmic `web'.  Toward small scales, the number, $N$, of collapsed CDM halos increases exponentially as mass decreases, $\mathrm{d}N/\mathrm{d}M_{\rm vir}\propto M_{\rm vir}^{-1.9}$ \citep{springel08}, where $M_{\rm vir}$ is the `virial' or equilibrium mass of the halo.  The internal structure of CDM halos is `universal' in the sense that halos on all scales approximately follow a mass-density profile, $\rho(r)=\rho_s(r/r_s)^{-1}[1+(r/r_s)]^{-2}$ \citep{navarro96}, specified by two scaling parameters whose correlation reflects dependence of halo `concentration'---the ratio of a halo's virial radius to its scale radius $r_s$---on the Universe's mean density at the time of collapse \citep{navarro96}.

\smallskip
So long as dark matter goes unobserved, tests of these predictions require inferences based on observations of luminous structure.  An obvious strategy is to consider galaxies as tracers of the abundance and spatial distribution of dark matter halos.  Indeed observational surveys reveal a luminous cosmic web that bears striking resemblance to the simulated dark one \citep{geller89,springel06}, indicating good agreement between CDM simulations and observations on large scales.  More quantitative comparisons on smaller scales often involve schemes for mapping of galaxies with specific observable properties onto halos with particular simulated properties.  The simplest assume that galactic luminosity is a monotonic function of halo mass, enabling unique association of observed galaxies with simulated halos \citep{conroy06}.  Such mapping onto a simulated halo population can translate one macroscopic observable (say, the luminosity function, $\deriv N/\deriv L$, of galaxies) into reliable predictions about another (the spatial distributions of those same galaxies).  Indeed it is now common to use such schemes to study galaxy formation and evolution under the assumption that CDM cosmological simulations accurately portray the underlying dark matter structure.

\section{Problems at the Smallest Galactic Scales}
\label{sec:smallscales}

However, several tantalizing hints to the contrary have come from studies of dwarf galaxies, the smallest objects associated empirically with dark matter clustering.  Dwarf galaxies are extreme not only in terms of size---they also include the most dense, least luminous, oldest and even the nearest (the Milky Way hosts $\sim 25$ known dwarf satellites) galaxies known.  There is ongoing and vigorous debate about what the number and internal structure of these dwarfs might be telling us about a minimum scale for dark matter clustering.  Here we discuss four apparent discrepancies between observations and CDM simulations on the smallest galactic scales.

\subsection{Accounting}
\begin{figure}
  \begin{tabular}{ll}
  \includegraphics[height=3in,width=3in]{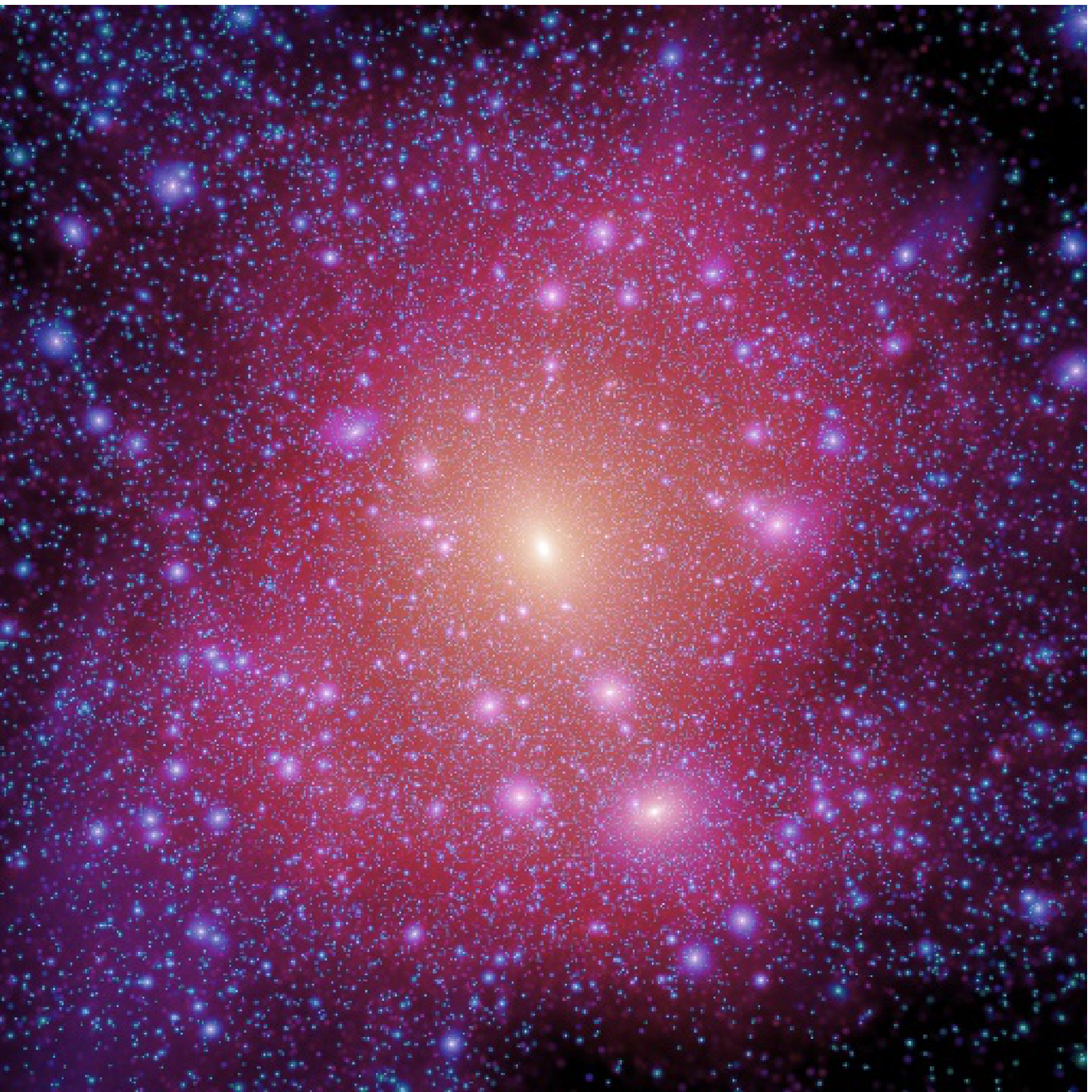}&\includegraphics[height=3in,width=3in]{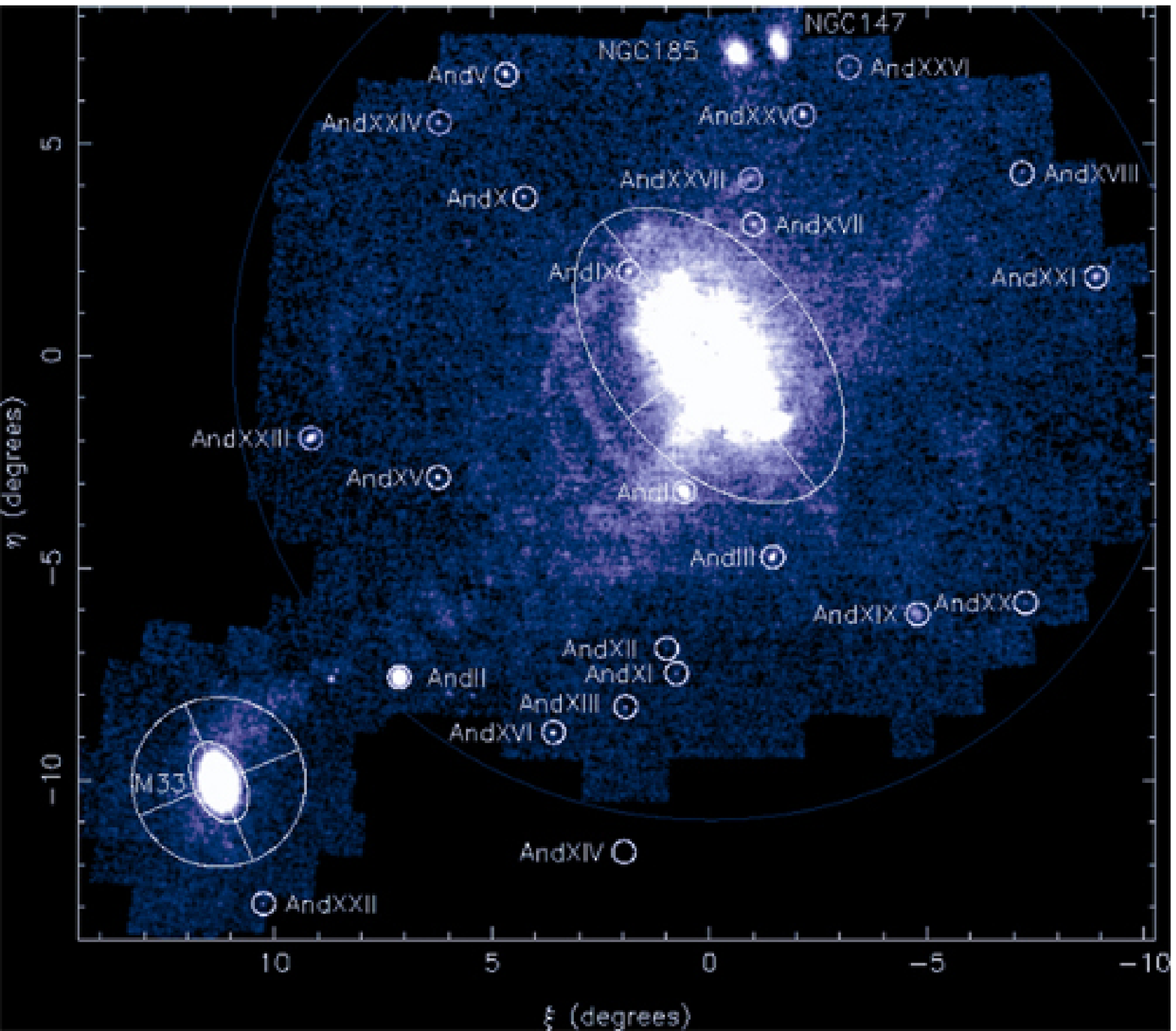}\\
  \end{tabular}
  \caption{Cold dark matter's `missing satellites' problem.  \textbf{Left:} spatial distribution of dark matter (pixel brightness scales with density) obtained in the $\sim 1$ Mpc$^2$ simulation box from the \textit{Aquarius} run, a high-resolution, dissipationless $\Lambda$CDM simulation by V.\ Springel, et al., originally published in Monthly Notices of the Royal Astronomical Society (2008, Vol.\ 391---reproduced with permission).  \textbf{Right:} spatial distribution of luminous stars observed in a deep, wide-field ($0.4$ Mpc $\times 0.4$ Mpc) survey of the Andromeda galaxy, originally published by J.\ Richardson, et al., in The Astrophysical Journal (2011, Vol.\ 732---\copyright AAS, reproduced with permission); Andromeda's stellar disc lies within the bright central region \cite{richardson11}.  The simulated dark matter distribution shows much richer `substructure' than does the observed luminous distribution.  }
  \label{fig:missingsatellites}
\end{figure}

As discussed above, simulations indicate that the smallest dark matter `micro-halos' collapse first, then merge and accrete nearby neighbors to form larger `subhalos'.  Subhalos then merge and accrete other subhalos and surviving microhalos, eventually forming halos sufficiently massive to host galaxies like the Milky Way.  However, the growth of such a massive halo does not entirely consume all of the smaller subhalos and even-smaller microhalos in its vicinity, many of which should survive as satellites (and satellites of satellites, etc.).  Thus CDM simulations unambiguously predict that the Milky Way's dark matter halo should be surrounded and orbited by the surviving siblings of its own building blocks (Figure \ref{fig:missingsatellites}, left panel).  

\smallskip
Indeed the Milky Way is surrounded by dwarf galaxies, most of which are near-spherical blobs called `dwarf spheroidals' (dSphs), whose own internal kinematics suggest they are hosted by dark matter subhalos.  These dwarfs give the opportunity to study relics of the Milky Way's formation and to test the CDM scenario.   When comparing CDM simulations to the observed dwarf galaxies, the most obvious discrepancy is a difference in numbers.   The left-hand panel of Figure \ref{fig:missingsatellites} shows the spatial distribution of dark matter as realised in a recent, high-resolution, dissipationless $\Lambda$CDM simulation of the formation of a Milky Way-like halo \citep{springel08}.  The right-hand panel is generated from real observational data and shows the spatial distribution of luminous stars observed in a recent survey of the Andromeda galaxy \cite{mcconnachie09}, which is similar to the Milky Way in terms of size, luminosity and mass and---unlike the Milky Way---offers Earthlings an external view.  This side-by-side comparison reveals that the simulated dark matter distribution shows much richer `substructure' (i.e., the thousands of small, dense `subhalos' surrounding the dominant central halo) than does the observed luminous distribution, which reveals several tens of dwarf-galactic satellites.  While simulations generally produce hundreds to thousands of satellite subhalos capable of binding stars orbiting at the speeds observed in dwarfs ($\sim 10$ km s$^{-1}$), the known dwarf-galactic satellites of the Milky Way and Andromeda number only in the tens (Figure \ref{fig:missingsatellites}, right panel).  

\smallskip
This deficit, known as the `missing satellites' problem, has persisted since the late 1990s, when CDM simulations first resolved Milky Way-like halos \citep[][]{klypin99}.  At that time the observational census included only $\sim 10$ known satellites of the Milky Way.  Since then, discoveries of $\sim 15$ `ultrafaint' ($10^3\simless L_V/L_{V,\odot}\simless 10^5$) Galactic satellites \citep[][]{belokurov07} have more than doubled that number, but have not closed the gap \citep{koposov08}.  Most recently, surveys with radio telescopes sensitive to emission from neutral hydrogen indicate that the mismatch extends even to isolated dwarf galaxies, for which the observed velocity function, $\mathrm{d}N/\mathrm{d}V$, begins to deviate from the simulated one at circular velocities as large as $V\sim 100$ km s$^{-1}$ ($M_{\rm vir}\sim 10^{11} M_{\odot}$; \cite{papastergis11}).

\smallskip
It has long been recognised that this kind of accounting problem might be explained within the CDM framework if star formation is sufficiently suppressed in low-mass CDM subhalos, such that most never exceed observational thresholds for detection as luminous galaxies.  Plausible scenarios invoke inefficient cooling of star-forming gas in subhalos of $M_{\rm vir}\simless 10^{8} M_{\odot}$), and/or loss of baryons to the Milky Way's external gravitational field \citep[][and references therein]{kravtsov10}.  Regardless of what particular mechanism is responsible, comparisons of CDM subhalo populations to the luminosity function of Milky Way satellites generally require that the stellar mass formed (and retained) per unit halo mass decreases sharply toward low masses.  It then becomes feasible to associate the observed satellites with only the most massive simulated subhalos \citep{stoehr02} and to suppose that the Galactic halo is teeming with less-massive and near-completely dark CDM subhalos.  

\subsection{Normalisation}
\label{subsec:tbtf}
\begin{figure}
  \begin{tabular}{ll}
    \hspace{-0.7in}\includegraphics[height=4in]{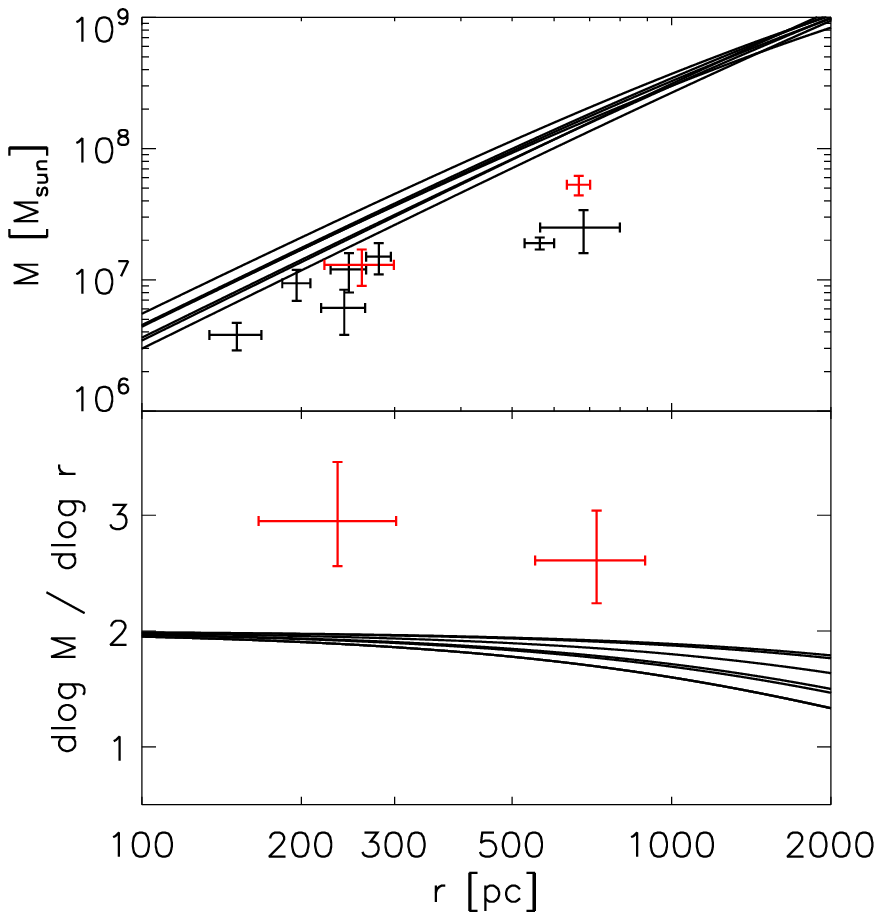}&\hspace{-0.6in}\includegraphics[height=3in]{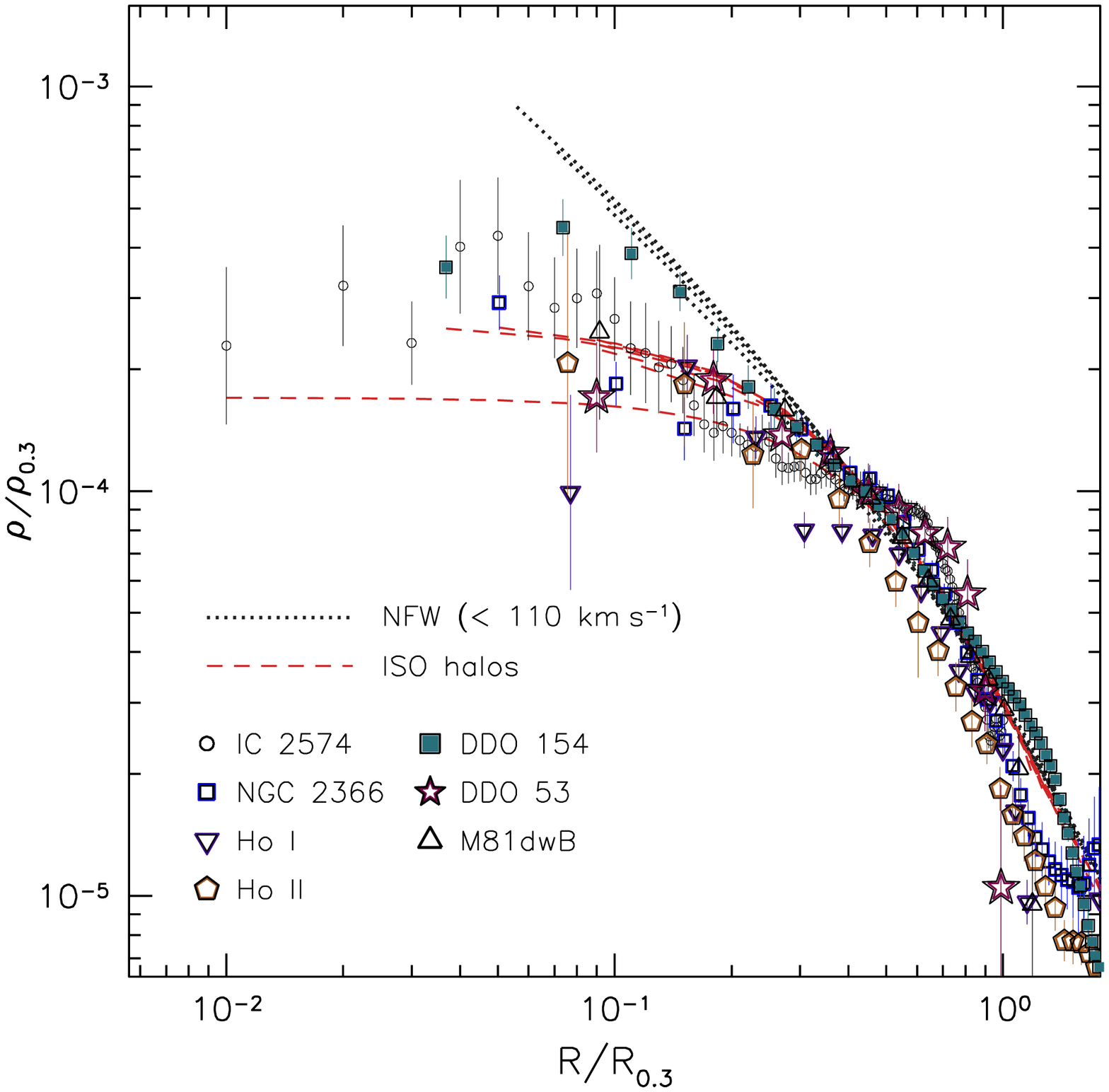}\\
  \end{tabular}
  \caption{`Core/Cusp' and `Too-Big-to-Fail' problems.  \textbf{Left:} Black curves display profiles of enclosed mass, $M(r)$ (top), and the logarithmic slope thereof (bottom), for the most massive CDM subhalos formed in dissipationless simulations of the Milky Way neighborhood \citep{springel08}.  Data points indicate estimates of 1) dynamical masses enclosed within projected halflight radii of the Milky Way's most luminous dwarf spheroidal satellites \citep[][top panel]{walker09d} and, for two dwarfs (shown in red) with estimated core sizes, 2) the slopes $\Delta \log M/\Delta\log r$ defined by masses enclosed within the different halflight radii of distinct stellar subpopulations \citep[][bottom panel]{wp11}.  With respect to observations of real dwarf satellites, the simulated subhalos have more mass enclosed at the measured halflight radii (Section \ref{subsec:tbtf}) and their enclosed-mass profiles have shallower slopes (Section \ref{subsec:slope}).  \textbf{Right:} Dark matter halo density profiles inferred from neutral-hydrogen rotation curves of seven low surface brightness galaxies, originally published by Oh et al.\ in \textit{The Astronomical Journal} (2011, Volume 141---reproduced by permission of the AAS).  The dotted black curve indicates the `cuspy' density profile that diverges toward small radii and characterises CDM halos formed in dissipationless N-body simulations.  Profiles inferred from observations of real galaxies (dotted red curves) tend to be `cored',  converging to approximately constant central density.  }
  \label{fig:tbtf}
\end{figure}

Undermining such a solution, however, are further discrepancies that arise when comparing internal structural properties of simulated subhalos to those of the subhalos inferred from observations.  From simple dynamical arguments, the characteristic, or `effective' radii\footnote{By convention, $R_e$ is the radius of the circle that encloses half of an object's light as seen in projection on the sky.}, $R_e$, and internal velocity dispersions\footnote{`Velocity dispersion' is the standard deviation of the velocity distribution.  For elliptical and spheroidal galaxies that are supported against gravity by random motions rather than ordered rotation, the characteristic orbital speed is given by the velocity dispersion rather than by rotational velocity.}, $\sigma$, measured for dSph stellar populations translate into upper limits on the masses, $M(R_e)$, enclosed within $R_e$.  The upper left-hand panel of Figure \ref{fig:tbtf} shows upper limits inferred for the Milky Way's eight most luminous dSphs; these values are all systematically smaller---by factors of $\sim 2-3$---than enclosed masses at similar radii in the $\sim 10$ most massive subhalos produced in published, high-resolution simulations of Milky-Way-like environments \citep{boylan-kolchin12}.  So it seems that the most massive simulated subhalos cannot host the most luminous dSphs after all.  If simulations have rendered the Milky Way's subhalo population accurately, then the most luminous dSphs must somehow live in some subset of less-massive subhalos while the most massive subhalos remain undetectably dark.  This implausibly fine tuning has been dubbed the `Too Big to Fail' (TBTF) problem \citep{boylan-kolchin11}.     

\smallskip 
Insofar as one should expect the most massive subhalos to host the most luminous satellites, TBTF amounts to a normalisation problem.  In principle it could be resolved with downward revision of simulated subhalo masses, as might be justified if the Milky Way's halo is less massive than presumed ($M_{\rm vir}\sim 10^{12}M_{\odot}$) when selecting analogs from simulations\footnote{The simulations that have been used to formulate the TBTF problem select halos of mass $M_{\rm vir}\sim 10^{12}M_{\odot}$ to represent the Milky Way.  In general the mass of a CDM halo is correlated with the masses of its most massive subhalos, such that if the halo of the real Milky Way is less massive than has been presumed, then its most massive subhalos will also be less massive, perhaps sufficiently so to remove the TBTF problem \citep{vera-ciro13}.   Undermining this possibility are the relatively large distance ($D\sim 250$ kpc) and Galactocentric radial velocity ($v\sim 175$ km s$^{-1}$) measured for the Leo I dSph, which have long implied a large Milky Way mass if Leo I is bound \citep{zaritsky89}.  Incorporating recent measurements of Leo I's motion, comparisons to CDM simulations indicate Leo I is likely to be bound if the Milky Way has virial mass $M_{\rm vir}\simgreat 10^{12}M_{\odot}$ \citep{boylan-kolchin13}.}, and/or if cosmic variance is sufficient to explain deviations from predictions derived specifically from the small number of simulations considered.  But the simplest formulation of the TBTF problem---\textit{the mean densities inferred within the central few-hundred pc of the most luminous dSphs are systematically smaller than predicted in CDM simulations}---evokes a further discrepancy that can be attributed neither to cosmic variance nor to uncertainty in the Milky Way's mass.

\subsection{Slope}
\label{subsec:slope}
Specifically, mass-density profiles inferred for galactic dark matter halos tend to have inner slopes that deviate systematically from the CDM predictions discussed in Section \ref{sec:scales}.  Instead of the centrally-divergent `cusps'---$\lim_{r\rightarrow 0}\rho(r)\propto r^{-1}$---that characterise inner regions of simulated halos \citep{navarro96}, dynamical mass profiles derived from stellar- and gas-dynamical data of dwarf galaxies tend to favor large ($\sim 1$ kpc) central `cores' of near-uniform density \citep[][and references therein]{deblok10}.  This `core/cusp' problem has a long history of contention.  Early reports of large, homogeneous dark matter cores, inferred from hydrogen rotation curves of gas-rich dwarf galaxies and low-surface-brightness galaxies, were scrutinised for biases that might arise from inadequate resolution, centering errors and/or non-circular motions.  More recently, the availability of high-resolution, two-dimensional velocity fields has largely mitigated such concerns.  While cusped density profiles can be fit in some cases, cored profiles are favored by the vast majority of dark-matter-dominated galactic rotation curves \citep[see, e.g., Figure 7 of reference ][]{oh11}.  

\smallskip
Battle lines have shifted recently toward pressure-supported systems, where dynamical inferences about mass distributions must rely on statistical arguments.  As a result, conclusions vary widely.  For example, reported detections of large dark matter cores in at least some of the Milky Way's dSph satellites \citep{wp11,amorisco12,jardel12} are contested by other analyses---often using the same data---that find consistency with cusped profiles \citep{strigari10,breddels13a,jardel13}.  Such disagreements arise from the application of qualitatively different kinds of analysis and inference \citep[for recent reviews, see references ][]{walker13,strigari13,battaglia13}, which obviously vary in terms of reliability and sensitivity to features in the data.  Since it is difficult to say \textit{a priori} which method is most appropriate in a given situation, the current state of disharmony demands participation in community-wide `data challenges'.  That is, dynamical modelers must compare the relative biases and sensitivities of various methods by applying them to common sets of artificial data sets that include realistic violations of idealised modeling assumptions\footnote{The first such comparisons are underway, having commenced with the \textit{Gaia Challenge} workshop held August 2013 in Surrey, UK; see http://astrowiki.ph.surrey.ac.uk/dokuwiki/doku.php.}.  Stakes are high, as the presence of dark matter cores in the Milky Way's dwarf satellites would mean that all three problems discussed above occur on the same scale, placing multiple and perhaps irreconcilable demands on possible solutions \citep{penarrubia12}.

\smallskip
At the upper end of the mass spectrum, inferences about inner density profiles of dynamically relaxed galaxy clusters benefit from availability of independent dynamical probes: stellar and galactic kinematics, X-ray temperature maps and gravitational lensing.  Given this leverage, the primary uncertainty is not necessarily inference of the central dynamical mass distribution, but is often domination of the dynamical mass by the contribution from stars in the most luminous, central galaxy.  Estimation of the dark matter density profile then requires careful subtraction of that galaxy's stellar mass profile, which itself is measured only indirectly.  Nevertheless, such estimates have become feasible given rapid progress in modeling of stellar populations \citep{conroy10b}.  Recent estimates that combine constraints from resolved stellar kinematics, strong and weak lensing suggest that while the \textit{total} density profiles of relaxed clusters are consistent with cusped CDM halos, the subdominant contributions from dark matter are significantly shallower near halo centers \citep[][]{newman13}, consistent with cores of size $\sim 10$ kpc \citep{newman13b}.  

\subsection{Discs of Satellites?}
\label{subsec:dos}
Finally, there is mounting evidence that the Milky Way's and Andromeda's dwarf satellite populations do not have the isotropic spatial configuration that one might, perhaps naively, expect for CDM subhalos.  In particular, a recent analysis of the positions and line-of-sight velocities of Andromeda's $\sim 30$ known dSphs concludes that approximately half are co-rotating within a plane of diameter $\sim 400$ kpc and thickness $\sim 20$ kpc \citep{ibata13}.  Although limited by the incomplete coverage of available sky surveys, there is evidence for similar anisotropy among the Milky Way's satellites \citep{metz07}.  These discoveries have generated ongoing debate about whether such configurations are feasible within the CDM framework---perhaps the result of subhalo accretion along filaments and/or in groups \citep{libeskind05}---or whether instead they imply that the satellites formed from gas ejected during tidal disruption events \citep{pawlowski13b}.  Like the core-cusp problem, this issue is largely unsettled.  Further progress requires quantification of the rarity of such alignments in real and simulated Universes, with 1) observations of low-luminosity satellite populations around more Milky-Way-like galaxies, and 2) high-resolution simulations of large numbers of Milky-Way-like galaxies.

\section{Complexity}
\label{sec:complicated}

The missing-satellites, TBTF and core/cusp problems have motivated suggestions that the dark matter model may require additional complexity that would naturally give rise to the small-scale clustering limits inferred for real galaxies.  For example, particles of sufficiently low mass (bounded at $m_{\chi}\simgreat 3$ keV by clustering observed in hydrogen absorption lines at high redshift \cite{viel13}) can decouple from photons at near-relativistic velocity dispersions in the early Universe, letting them travel relatively large distances before falling into gravitational potential wells.  Models based on such `warm' dark matter (WDM) introduce minimum clustering scales in terms of length and mass, plausibly solving the missing-satellites problem by truncating the halo mass function on the same scales as the observed galaxy luminosity function \citep{bode01}.  Alternatively, particles that interact with each other via a `Yukawa-like' coupling \citep{loeb11} can efficiently self-scatter out of high-density regions.  Such `self-interacting' dark matter (SIDM) models can naturally flatten density profiles near the centers of individual dark matter subhalos, thereby giving plausible solutions to the core/cusp and TBTF problems \citep{spergel00}.  Interestingly, neither WDM nor SIDM alone seem able to solve all of CDM's small-scale problems simultaneously: viable WDM models would suppress formation of the smallest subhalos without appreciably altering the internal structure of individual subhalos \citep{maccio13}; conversely, viable SIDM models would alter the internal structure of individual subhalos without significantly suppressing formation of the smallest ones \citep{zavala13}.  It is therefore tempting to propose even further complexity of the dark matter model by invoking a mixture of CDM, WDM and/or SIDM components \citep[][]{fan13a}.  

\smallskip
Even so, there remain viable solutions that invoke greater complexity without putting it into the dark matter model.  The small-scale problems in Section \ref{sec:smallscales} seem to indicate a genuine problem with the basic assumption underpinning all dissipationless CDM cosmological simulations---namely, that structure formation is dominated by gravitational interactions among standard CDM particles.  Maybe the problem is indeed CDM, but concluding so would require first knowing that CDM clustering is unaffected by the baryon-driven processes that get neglected (by definition) in dissipationless N-body simulations---i.e., that gas cooling and inflows, disk formation, stellar winds, radiation pressure, supernova explosions and subsequent outflows all have negligible impact on the abundance and internal structure of low-mass dark matter subhalos.  

\smallskip
Many recent hydrodynamical simulations suggest the opposite.  For example, CDM halos contract as cooling gas slowly sinks toward the center \citep[][]{blumenthal86}.  More importantly for low-mass subhalos, if sufficiently strong winds expel gas rapidly---and, for greater effect, repeatedly---from the center of a CDM subhalo, the result can be a gradual flattening of the central cusp\footnote{The physical mechanism underlying this transformation can be understood analytically as a form of `violent relaxation' \citep{lynden-bell67}, whereby the energy released from baryon-physical mechanisms---e.g., supernova explosions and subsequent blowout of interstellar gas---is transferred irreversibly to the orbits of dark matter particles via rapid fluctuations in the gravitational potential \citep{pontzen12}.} \citep[e.g.,][]{pontzen12}.    Lowered binding energies would then render newly-cored CDM subhalos more susceptible to tidal disruption in the vicinity of a parent halo, offering a handy mechanism for reducing the abundance of galaxy-hosting CDM subhalos around a Milky Way-like galaxy \citep{brooks13}.   Thus the inclusion of realistic baryon physics in cosmological simulations might modify CDM `predictions' of both the abundance \textit{and} internal structure of low-mass galaxies, raising the possibility of preserving CDM's simplicity by shifting needed complexity from the dark matter model onto the broader model for galaxy formation.  Indeed state-of-the-art codes for incorporating baryon physics into N-body simulations are currently being used systematically to explore effects of varying the many free parameters available for implementing processes associated with star formation, winds and active galactic nuclei \citep{vogelsberger13}.  Ultimately, testability of the CDM hypothesis will depend on the ability of such simulations to elicit robust predictions for visible galactic structure.  

\section{Simplicity}
However complicated galaxy formation may be, the outcome is a universe in which the sizes and internal velocities of galaxies scale simply and predictably with baryon content.   For example, the maximum rotational velocities of spiral galaxies correlate strongly with luminosity, $L$ \citep{tully77}, a relation that becomes even tighter if luminosity is replaced with baryonic mass \citep{mcgaugh00}.  In the 3-D space defined by axes of size, $R$, surface brightness, $\Sigma=L/(2\pi R^2)$, and velocity dispersion, $\sigma$, pressure-supported galaxies populate a `fundamental plane'---$R\propto \sigma^{\alpha}\Sigma^{\beta}$, with $\alpha\sim 1.5$ and $\beta\sim -0.75$ \citep{dressler87}.  `Tilt' with respect to the virial relation derived from the assumption that $\sigma^2\propto L/R$ (giving $\alpha=2$, $\beta=-1$) correlates with observables, thereby enabling empirical calibration of a `fundamental manifold' relation that can accommodate \textit{all} spheroids ranging from globular clusters to galaxy clusters \citep{zaritsky11}.  
\label{sec:simplicity}

\begin{figure}[ht]
\includegraphics[width=15cm]{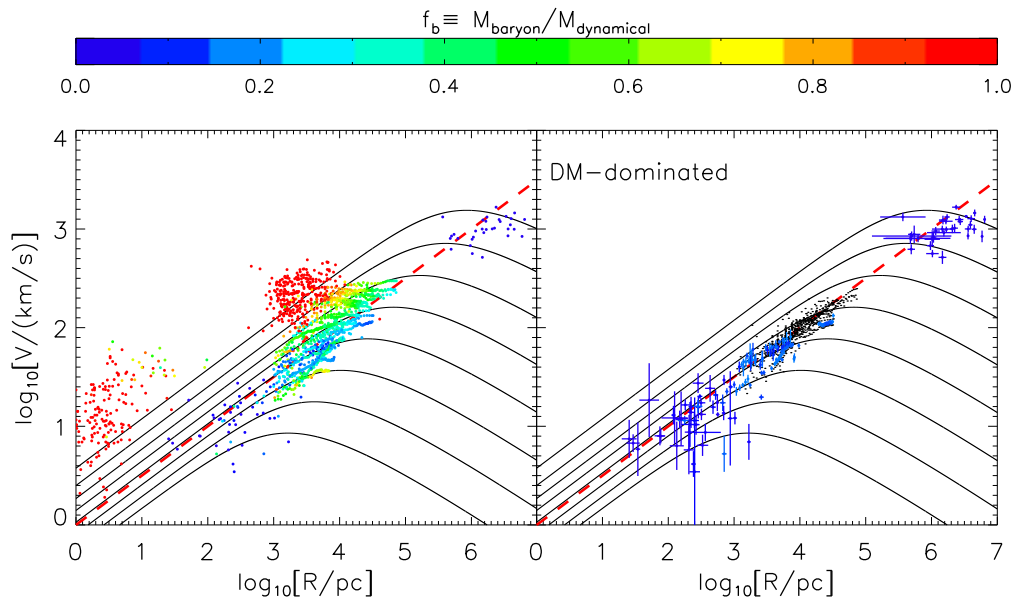}
\caption{Orbital velocity, $V$, versus radius, $R$, for gravitationally bound objects ranging from globular star clusters to galaxy clusters.  Colour indicates the fraction of mass contributed by baryons, $f_{\rm b}\equiv M_{\rm b}/M_{\rm dyn}$.  The right-hand panel isolates systems with dynamics dominated by dark matter ($f_{\rm b} \le 0.15$; for clarity, error bars are included only in the right-hand panels).  Solid black lines indicate circular-velocity curves calculated for `NFW' \cite{navarro96} profiles, representing  CDM halos with virial masses $\log_{10}[M_{vir}/M_{\odot}]=\{8,9,...,15\}$ and mass/concentration adopted from the \textit{MultiDark} and \textit{Bolshoi} cosmological simulations \cite{prada12}.  Small black points in right-hand panel indicate contributions of dark matter to resolved rotation curves of spiral galaxies \cite{mcgaugh07} (i.e., after subtracting estimated contributions of stars and gas; \cite{mcgaugh07}).  Red dashes mark a line of constant acceleration, $a_{\rm DM}\sim 1$ km$^{2}$s$^{-2}$pc$^{-1}$, about which the blue points scatter with rms deviation $\sim 0.2$ dex.}
\label{fig:dm}
\end{figure}

\smallskip
Such correlations of dynamical and baryonic properties carry information about the astrophysics of galaxy formation, of which dark matter is but one component.  In the interest of isolating dark matter phenomenology, Figure \ref{fig:dm} depicts the relationship between purely \textit{dynamical} components of galactic scaling relations, using data we have gathered from the literature (see Appendix for details).  
For rotationally-supported galaxies, Figure \ref{fig:dm} plots measurements of orbital velocity, $V$, versus orbital radius, $R$, including resolved stellar and cold-gas rotation curves containing between $1$ - $70$ independent measurements per galaxy.  For pressure-supported spheroids, the plotted quantities are velocity dispersion and effective radius (the radius of the circle that encloses half of the tracers as projected on the sky).  Colours indicate estimates of baryon fractions, $f_{\rm b}\equiv M_{\rm b}/M_{\rm dyn}$, where the baryonic mass, $M_b$, can include contributions from stellar and gaseous components, and the dynamical mass is $M_{\rm dyn}\equiv RV^2/G$.  

\smallskip
Larger systems tend to have larger internal velocities, but for a given size, velocity increases systematically with baryon fraction.  These dependences are such that if we focus on the dark-matter-dominated regime and consider only data points with baryon fractions smaller than the cosmic one---i.e., $f_{\rm b}\simless 0.15$, thereby excluding all globular clusters and luminous ellipticals and including most dwarf galaxies, low-surface-brightness galaxies, outer regions of spiral galaxies, and galaxy clusters---we recover the scaling $V^2\propto R$, about which rms deviation is $\sim 0.2$ dex.  This dependence can be represented by a single quantity that has units of acceleration: $a_{\rm DM}\equiv V^2/R\sim 1$ km$^{2}$ s$^{-2}$ pc$^{-1}\approx 3\times 10^{-9}$ cm s$^{-2}\approx 0.3c\sqrt{G\rho_{\rm DM,0}}\approx 0.2 c\sqrt{G\rho_{\rm v}}$ (dashed red line in Figure \ref{fig:dm}).  

\smallskip 
The same scaling has previously been identified for the contribution of dark matter to the rotation curves at intermediate radii of spiral galaxies (small black points in right-hand panels of figure \ref{fig:dm}; \cite{mcgaugh07}), and subsequently extended to include Milky Way dSphs \citep{walker10}, thereby generalising earlier claims about common dark matter properties among dSphs \citep{mateo93,strigari08,walker09d}.  More recently, a kinematic study of Andromeda's dSph satellites finds an offset toward smaller velocity dispersion at fixed size \citep{collins13b}; however, this offset is driven primarily by three (from a population of 25) low outliers (visible in Figure \ref{fig:dm} at $\log_{10}[R/\mathrm{pc}]\sim 3$) that are suspected of having migrated downward due to tidal disruption \citep{collins13b}.  Figure \ref{fig:dm} suggests that the relation stretches the other way too, to include galaxy clusters (see Appendix \ref{sec:galaxyclusters}) and thereby to span $\sim 5$ orders of magnitude in size.

\smallskip
A `characteristic acceleration' can incorporate previous conclusions that galactic dark matter halos have a characteristic surface density.  Specifically, for spiral and dSph galaxies spanning $\sim 5$ orders of magnitude in luminosity, fits of cored halo profiles to rotation curves and velocity dispersion profiles reveal an anti-correlation among core radius, $r_c$, and core density, $\rho_c$, such that inferred surface densities maintain a common value: $\Sigma_{\rm DM}\equiv r_c\rho_c \sim 150 M_{\odot}\mathrm{pc}^{-2}$ \citep{kormendy09}.  Via the conversion $\Sigma_{\rm DM}\sim a_{\rm DM}/G$, such a characteristic surface density is equivalent to a characteristic acceleration.  However, while acceleration is a straightforward combination of observables, surface density refers to a mass distribution that must, in some intermediate step, be inferred from those observables under a particular dynamical model.  Thus the alternative characterisation in terms of acceleration provides a more economical and robust representation of what is likely the same phenomenon.

\smallskip
There is already an acceleration scale known to be associated with dark matter phenomenology.  Galactic rotation curves reveal that dynamical mass systematically exceeds baryonic mass, $M_b$, at galactic radii where $GM_{b}/R^2\simless a_0$, with $a_0\sim 10^{-8}$ cm s$^{-2}$ \citep{mcgaugh04}.  This is the behavior that motivates the alternative paradigm of Modified Newtonian Dynamics (MOND), under which the acceleration of a test particle becomes $g=\sqrt{GM_ba_0/R^2}$ in the limit that $GM_b/R^2\ll a_0$ \citep{milgrom83}, thereby describing the motions of outer particles without invoking dark matter as a substance.
Interestingly, MOND implies that the dark matter halos inferred under Newtonian gravity have a maximum internal acceleration, $a_{\rm max}\sim 0.3a_0$ \citep{brada99}, which takes the same value as the acceleration we have called $a_{\rm DM}$.  In the right-hand panel of Figure \ref{fig:dm}, the only (marginally) significant outlier with $V^2/R>a_{\rm DM}$ is the galaxy cluster Abell 957 \citep{rines06}, with $\log_{10}[V^2/R/(\mathrm{km}^2\mathrm{s}^{-2}\mathrm{pc}^{-1})]=0.6 \pm 0.3$.  Thus MOND predicts not only an acceleration scale associated with baryonic mass, but also a scale for the acceleration attributed to dark matter halos in the Newtonian framework.   Galaxies appear to respect these predictions.

\smallskip
On the other hand, while the CDM framework can accommodate the appearance of an acceleration scale in Figure \ref{fig:dm}, it does not clearly predict one.  Solid curves in Figure \ref{fig:dm} represent circular velocity profiles, $v_{c}=\sqrt{GM(r)/r}$, for CDM halos of virial mass $\log[M_{\rm vir}/M_{\odot}]=8,9,...,15$ (the range of halo masses associated with objects ranging from dwarf galaxies to galaxy clusters), calculated using analytic functions fit to halos produced in dissipationless simulations \citep{prada12}.  At all radii, higher-mass halos have larger circular velocities than do lower-mass halos.  Furthermore the curves in higher-mass halos rise over a larger fraction of their total volume than do the curves in lower-mass halos, a consequence of the tendency for higher-mass halos to be less centrally concentrated.  As a result, the full set of curves spans more than an order of magnitude in velocity at a given size.  Yet the observational data populate only a small portion of this available range, approximately tracing a straight line through the set of bending curves.  While it is always possible to invoke a CDM halo having the necessary mass and concentration to go through a given data point, and thereby to fit all the data by invoking a suitably large range of CDM halos, one can fit the data more simply just by drawing a straight line of constant acceleration.  

\smallskip
The ability of $\Lambda$CDM to explain this apparent simplicity of galactic structure is largely to be determined, pending outcomes of current and future efforts focused on identifying, understanding and simulating relevant baryon-physical processes \citep{vogelsberger13}.  Many investigators have already shown, working within a $\Lambda$CDM context, that particular sets of assumptions about and/or treatments of dissipative processes can reproduce the general trends encapsulated by galactic scaling relations \cite{vandenbosch00}; the question is now whether simulations that include the necessary complexities of baryon physics can account for the small scatter about the observed relations.  The rapid progress of simulations strongly motivates efforts to settle observational controversies about internal structure of the smallest dwarf galaxies (Section \ref{subsec:slope}), which impose a boundary condition on the baryon physics of galaxy formation and will likely determine just how much complexity is required.  Then the ultimate astrophysical test of CDM will be whether future simulations can reproduce galactic scaling relations while \textit{simultaneously} giving accurate predictions regarding the abundances and internal structures observed for the smallest and least luminous dwarf galaxies.

\section{Is the Universe Simpler than $\Lambda$CDM?}
\label{sec:question}
We have examined the $\Lambda$CDM cosmological model in terms of its simplicity.  The model combines the simplest substance-based interpretations of cosmic acceleration and galactic dynamics.  It is calculable, enabling robust predictions about the growth of structure in a universe dominated by vacuum energy and cold dark matter.  Thus far its simplicity is largely supported by agreement with observations on perhaps all but the smallest ($\simless 10$ kpc) cosmic scales, where alternative particle models are competing with CDM to explain galactic structure.  The CDM version seems to require that structure formation is \textit{not} entirely dominated by dark components, but rather is subject to the complexities introduced by the dissipative, explosive and difficult-to-simulate physics of baryons.  

\smallskip
On the same scales where CDM must invoke this complexity, galaxies exhibit simplicity.  Their dynamical properties relate to their luminous properties such that the most reliable predictions often come not from simulations, but from other observations.  For the entire range of gravitational potentials dominated by dark matter, a `characteristic acceleration' seems to encode a relationship between orbital size and velocity, the two observables used to infer the existence of galactic dark matter.  Its value is $a_{\rm DM}\approx 0.3c\sqrt{G\rho_{\rm DM,0}} \approx 0.2c\sqrt{G\rho_v}$, where $\rho_{\rm DM,0}$ is the current value for the mean density attributed to dark matter in the Universe and $\rho_{\rm v}$ is the constant value for the mean density attributed to vacuum energy.  

\smallskip
In the current $\Lambda$CDM framework, the similarity of all of these quantities is coincidental.  Until $\Lambda$CDM can explain the apparent simplicity of galactic scaling relations it will be tempting to consider alternative models in which there may be an explicit  connection between the internal dynamics of dark matter halos and the Universe's overall composition.   \citep[e.g.,][]{bekenstein04,moffat06}.  Regarding the appearance of an acceleration scale for dark matter, future observations that extend Figure \ref{fig:dm} to higher redshift---particularly for the dwarf galaxies and galaxy clusters dominated by dark matter\footnote{Both types of object pose unique observational challenges.  The low luminosities of dwarf galaxies make them difficult to detect---and even more difficult to study spectroscopically---at high redshift.  The most massive and most luminous galaxy clusters are the latest-forming bound objects and are therefore least likely to be virialized at high redshift.}---might be able to test for a connection to the mean cosmic density of either dark matter, $\rho_{\rm DM}$, or vacuum, $\rho_{\rm v}$.  As the Universe expands, $\rho_{\rm DM}$ decreases as the inverse-cube of the expansion factor (i.e., mass is conserved), while $\rho_{\rm v}$ remains constant.  Therefore, if connected to the cosmic abundance of dark matter, $a_{\rm DM}$ will decrease with time and the value estimated from objects at redshifts $z\simgreat 1$ will be $\simgreat 10$ times larger than the value estimated from the sample of relatively nearby objects ($z\simless 0.1$) shown in Figure \ref{fig:dm}.  If connected to the density of vacuum energy, observations of objects at all redshifts should yield the same value of $a_{\rm DM}$ as the local sample.  In any case, a satisfactory model of the invisible must account for the structural regularities that we can see, and this remains the primary challenge facing $\Lambda$CDM.

\section*{Acknowledgements}
We thank Ken Rines for providing the SDSS-selected spectroscopic sample for CIRS galaxy clusters \cite{rines06}.  We thank Se-Heon Oh for providing the resolved rotation curve data presented by the THINGS survey \cite{oh11}.  We acknowledge helpful correspondence with Shantanu Desai.  MGW is supported by NSF grant AST-1313045.  AL is supported in part by NSF grant AST-1312034.

\section*{Notes on contributors}

Matthew G. Walker is assistant professor of Physics at Carnegie Mellon University in Pittsburgh, Pennsylvania.  Before arriving at CMU in 2013, Walker was a Postdoctoral Research Associate in the Institute of Astronomy at the University of Cambridge, England, and a Hubble Postdoctoral Fellow at Harvard College Observatory in Cambridge, Massachusetts.  An observational astronomer, Walker uses large telescopes to measure the motions of stars in nearby galaxies.

\smallskip
\noindent Abraham Loeb is the Frank B. Baird, Jr. Professor of Science at Harvard University in Cambridge, Massachusetts, where he also serves as chair of the Department of Astronomy and director of the Institute for Theory and Computation within the Center for Astrophysics.   Loeb has worked on a wide variety of topics in theoretical astrophysics, including but not limited to cosmology, black holes, and formation of the first galaxies.  He is an elected member of the American Academy of Arts \& Sciences.  

\appendices

\section{Data Adopted from the Literature}

Figure \ref{fig:dm} shows structural and kinematic data adopted from the literature and restricted to low redshift, $z\simless 0.1$.  Here we identify the adopted data sets and describe how we derive from each consistent estimates of characteristic radii, velocities and baryon fractions.   We estimate baryon fractions, $f_{\rm b}\equiv M_{\rm b}/M_{\rm dyn}$, using available estimates of baryonic masses (i.e., stellar plus gas masses) and define dynamical mass as $M_{\rm dyn}\equiv RV^2/G$.  For rotationally-supported spiral and low-surface-brightness galaxies, we adopt measurements of $R$ and $V$ directly from resolved rotation curves.  For pressure-supported systems, we equate $R$ with effective radius (radius of the circle that encloses half of the tracers as viewed in projection), and we equate $V$ with $\sqrt{3}\sigma$, where $\sigma$ is the line-of-sight velocity dispersion.  In cases where estimation of $f_b$ requires conversion of stellar luminosity to stellar mass, for simplicity we assume stellar mass-to-light ratios $\Upsilon_*=1$ in solar units in all bands.  In some cases, particularly for compact pressure-supported systems, the crudeness of these definitions results in unphysical estimates $f_{\rm b}>1$ (these objects are plotted in figure \ref{fig:dm} as though they have $f_b=1$); however, the affected objects (globular clusters, ultra-compact dwarf galaxies and luminous elliptical galaxies) are clearly baryon-dominated and therefore not relevant to conclusions about dark-matter dominated potentials.  

\subsection{Globular Clusters and Ultra-Compact Dwarf Galaxies}
\label{sec:glob/ucd}

For Galactic globular clusters, we adopt measurements of effective radii, $R_h$, V-band luminosities, $L_V$, and central stellar velocity dispersions, $\sigma_0$ from the catalog of W.\ Harris \cite{harris96} (2010 edition, including measurements of velocity dispersions \cite{pryor93,mclaughlin05}.  For M31 globular clusters, we adopt measurements of the same quantities from the survey published by J.\ Strader et al. \cite{strader11}  We supplement these data sets with additional data for compact stellar systems---globular clusters and ultra-compact dwarf galaxies (UCDs)---using the compilation of Mieske et al. \cite{mieske08}  Included objects are the large globular cluster G1 in M31 \cite{meylan01}, globular clusters in the giant elliptical galaxy NGC 5198 (Centaurus A, \cite{rejkuba07}), globular clusters and ultra-compact dwarf galaxies (UCDs) in the Virgo cluster \cite{hasegan05,evstigneeva07}, and UCDs in the Fornax cluster \cite{mieske08,hilker07}.  Figure \ref{fig:dm} plots $V=\sqrt{3}\sigma_0$ against $R=R_h$.   $0.5 \simless f_b \simless 70$, with median value of $3$.

\subsection{Dwarf Spheroidal Galaxies}
\label{sec:dsph}

For the 48 Local Group dwarf spheroidal (dSph) galaxies that have stellar-kinematic measurements, we adopt the halflight radii, $R_h$, V-band luminosities, and global velocity dispersions, $\sigma$, tabulated in the recent review by McConnachie \cite{mcconnachie12}.  For twelve of M31's dSph satellites (And I, III, V, VII, IX, X, XIII, XIV, XV, XVI, XVII, XXI), we replace the velocity dispersions with more recent measurements \cite{tollerud12,collins13}.  Figure \ref{fig:dm} plots $V=\sqrt{3}\sigma$ against $R=R_h$.  We estimate baryon masses as $M_{\rm b}=L_V\Upsilon_*$ assuming $\Upsilon_*=1$, which gives baryon fractions $3\times 10^{-4}\simless f_b\simless 0.6$, with median value $0.03$.  We obtain $f_{\rm b} < 0.15$ for 40 of the 48 dSphs.

For the Galactic dSphs Fornax and Sculptor, we include structural/kinematic parameters estimated for the overall stellar population as well as the parameters recently estimated for two chemo-dynamically distinct stellar sub-populations \cite{wp11}.  For both galaxies we estimate $f_b\sim 0.4$ for the inner sub-population and $f_b \simless 0.1$ for the outer sub-population.

\subsection{Low Surface Brightness Galaxies}
\label{sec:lsb}

We consider data for the nine low surface brightness galaxies (LSBs; UGC 4325, F563-V2, F563-1, DDO 64, F568-3, UGC 5750, F583-4 and F583-1) that have mass models presented in the second part of Table 3 from R. Kuzio de Naray et al.\ \cite{kuzio08}  Specifically, figure \ref{fig:dm} plots the same rotation curves plotted by Walker et al.\ \cite{walker10}.  These velocities have been corrected for pressure support (typically $\sim 8$ km s$^{-1}$), which generally has negligible effect.  The data include estimates of baryonic contributions, $V^2_{b}(R)=V^2_{HI}(R)+V^2_{*}(R)$, to the rotation curves.  Here, $V^2_{HI}(R)=GM_{HI}(R)/R$ is the rotation velocity due to the enclosed mass of neutral hydrogen and $V_{*}^2(R)=GM_{*}(R)/R=GL(R)\Upsilon_*/R$ is the rotation velocity due to the enclosed mass of stars, with $L(R)$ the enclosed stellar luminosity and $\Upsilon_*$ the stellar mass-to-light ratio (assumed to be inedependent of radius).  The mass of neutral hydrogen is estimated directly from the HI data.  The stellar luminosity is estimated from the (optical) surface brightness profile, and the stellar mass-to-light ratios are either adopted from previous work \cite{deblok98} or derived from stellar-population-synthesis models \cite{kuzio08}.  We use only data with $r\ge 1$ kpc because systematic uncertainties dominate deconvolution of rotation curves at smaller radii \cite{mcgaugh07}.  Taking $M_{\rm b}(R)=M_{HI}(R)+M_*(R)$, we estimate baryon fractions at each point in each rotation curve.  We obtain $0.05\simless f_b\simless 0.9$, with median value $0.25$.  We obtain $f_b<0.15$ for 36 of the 162 resolved measurements.

\smallskip
We also include data for seven LSBs (DDO 154, DDO 53, HoI, HoII, IC2574, M81dwB, NGC 2366) observed for the THINGS survey \cite{oh11} (rotation curve data were generously provided by Se Heon Oh, private communication).  These data include resolved rotation curves with $\sim 10-100$ independent measurements per galaxy.  As with the LSB sample of \cite{kuzio08}, we use only data with $r\ge 1$ kpc.  We obtain $0.09\simless f_{\rm b}\simless 0.8$, with median value $0.3$; 9 of 248 resolved measurements have $f_{\rm b}<0.15$.

\subsection{Spiral Galaxies}
\label{sec:spiral}

For spiral galaxies, we adopt the rotation curves published by McGaugh et al. \cite{mcgaugh07}\footnote{available at http://www.astroweb.case.edu/ssm/data/galaxy\_massmodels.dat}.  The data consist of HI rotation curves \cite{sanders02}, trimmed for quality to a formal accuracy of $5\%$ or better \cite{mcgaugh05}, and restricted to radii larger than $R\geq 1$ kpc because systematic uncertainties (e.g., noncircular motions, beam smearing) dominate the deconvolution of rotation curves at smaller radii.  The final sample contains $696$ independent, resolved rotation velocity measurements for 60 galaxies, spanning radii $1\simless R\simless 75$ kpc.  This sample covers virtually the entire range of spiral properties, including circular velocities $50 \simless V_f \simless 300$ km s$^{-1}$, disk scale radii $0.5 \simless r_d\simless 13$ kpc, baryonic masses $3\times 10^8 < M_b < 4\times 10^{11}M_{\odot}$, central surface brightnesses $19.6\simless \mu_{0,b}\simless 24.2$ mag arcsec$^{2}$, and gas fractions $0.07\simless f_g\simless 0.95$.  

\smallskip
The data for spirals include estimates of contributions to the rotation curves from stellar (bulge, disk) and HI components.  For the component due to stellar mass, we adopt the value of $\Upsilon_*$ that obtained by fitting the rotation curve with modified Newtonian dynamics (MOND  \cite{milgrom83}); these estimates of $\Upsilon_*$ generally stand in excellent agreement with population synthesis models \cite{bell01}.  Taking $M_{\rm b}(r)=M_{HI}(R)+M_*(R)$, we estimate baryon fractions ranging from $0.09 \simless f_{\rm b}\simless 0.97$, with median value $0.4$; 41 of 696 resolved measurements have $f_{\rm b}<0.15$.  In order to compare to results for objects that are intrinsically dominated by dark matter, black points in the right-hand panels of figure \ref{fig:dm} indicate estimates of the contribution of dark matter to these rotation curves \cite{mcgaugh07}: $V^2_{\rm DM}(R)=V^2(R)-V^2_*(R)-V^2_{HI}(R)$. 

\subsection{Elliptical Galaxies}
\label{sec:ellipticals}

We consider the stellar-kinematic data taken with the SAURON integrated-field spectrograph for the SAURON and ATLAS$^{\rm 3D}$ surveys \cite{emsellem04,cappellari13}, which provide a complete sample of 260 early-type galaxies out to a distance of $\sim 50$ Mpc.  We adopt tabulated values \cite{cappellari13} of galaxy distances, $r$-band effective radii $R_e$, $r$-band luminosities, $L_r$, and stellar mass-to-light ratios, $\Upsilon_*$, and velocity dispersions $\sigma_e$ measured interior to $R_e$.  Figure \ref{fig:dm} plots $R=R_e$ versus $V=\sqrt{3}\sigma_e$.  Estimating baryonic masses as $M_{\rm b}=M_*=L_r\Upsilon_*$, we obtain $0.9\simless f_{\rm b}\simless 3$, with median value $1.7$.  

\smallskip
The most useful tracers of gravitational potentials at large radii in giant elliptical galaxies tend not to be stars, but rather globular clusters and/or planetary nebulae (`PNe').  While both types of tracer enable estimates of scale radii and velocity dispersions, estimation of the relevant baryon fractions is complicated by uncertain stellar mass-to-light ratios and, in many cases, the additional contribution from ionized gas.  For the present study, we adopt kinematic data for PNe taken with the Planetary Nebula Spectrograph \cite{douglas02}.  Specifically, for six S0 galaxies we adopt $K$-band luminosities, $L_{\rm K,disk}$ and $L_{\rm K,sph}$, for stellar disk and spheroid components respectively, effective radii $R_e$ of spheroidal components, and  velocity dispersions, $\sigma_{\rm sph}$, estimated for the spheroidal component using velocity measurements for hundreds of individual PNe \cite{cortesi13b}.  The spatial distribution of PNe extends to several $R_e$ and approximately traces the spatial distribution of stars in the spheroidal component \cite{cortesi13}.  For each of the six galaxies, figure \ref{fig:dm} plots $R=R_e$ versus $V=\sigma_{\rm sph}$.  Assuming $K$-band stellar mass-to-light ratios of $\Upsilon_{\rm *,K}=1$ in solar units and estimating baryonic masses as $M_{\rm b}=M_*=(L_{\rm K,disk}+L_{\rm K,sph})\Upsilon_{*}$, we obtain $3\simless f_{\rm b}\simless 25$, with median value $7$.  

\smallskip
We add PNe kinematic data for another 6 early-type galaxies observed with the Planetary Nebula Spectrograph, along with previously published PNe kinematics for an additional 10 early-type galaxies compiled by Coccato et al. \cite{coccato09}.  We adopt $B$-band luminosities, $L_{\rm B}$, effective radii, $R_e$, and the RMS velocity, $V_{\rm rms}$, measured within $R_e$ from $\simgreat 100$ PNe in each galaxy.  For each of these 16 galaxies, figure \ref{fig:dm} plots $R=R_e$ versus $V=V_{\rm rms}$.  Estimating baryonic masses as $M_{\rm b}=M_*=(L_{\rm B}\Upsilon_{*}$ and assuming $\Upsilon_*=1$, we obtain $0.5\simless f_{\rm b}\simless 16$, with median value $0.7$.  

\subsection{Galaxy Clusters}
\label{sec:galaxyclusters}

We adopt SDSS spectroscopic velocities for thousands of galaxies projected along lines of sight to the 74 X-ray-selected (flux $f_{\rm X} \simgreat 3\times 10^{-12}$ erg s$^{-1}$ cm$^{-2}$) galaxy clusters at low redshift ($z\simless 0.1$), previously studied for the Cluster Infall Regions in the Sloan Digital Sky Survey (CIRS) sample \cite{rines06}.  For most of the CIRS clusters, the SDSS Main Galaxy Survey is $\sim 90\%$ complete to a spectroscopic limit of $r\sim 17.8$.  For each cluster, Rines et al. \cite{rines06} apply the `caustic' technique \cite{diaferio97,diaferio99} to projected phase-space data in order 1) to identify the cluster's center of gravity (finding a median difference of $109$ $h^{-1}$ kpc with respect to X-ray centers), 2) to estimate escape velocity as a function of radius, 3) to identify gravitationally bound members and 4) to recover cluster mass profiles at large (several Mpc) radius even when the usual assumptions of hydrostatic equilibrium are not valid.  It has previously been shown \cite{geller13} that these estimates agree well with independent measurements derived from weak lensing \cite{oguri10}.  

The homogeneity and spectroscopic completeness (to a useful limiting magnitude) of the SDSS catalog provides the rare opportunity to estimate directly a spatial scale for the galaxy population within a galaxy cluster.  For each cluster, we adopt estimates of line-of-sight velocity dispersion, $\sigma_g$,  from bound galactic members \cite{rines06}, as well as the projected radius to which the SDSS spectroscopic survey provides complete spatial coverage.  Using CIRS catalog, generously provided by Ken Rines (private communication), we identify 32 clusters for which all spectroscopically-confirmed, bound galactic members lie inside the radius of complete spatial coverage.  Considering only these clusters, we estimate the projected `half-galaxy' radius, $R_g$, as the median projected radius of bound galactic members.   Figure \ref{fig:dm} plots $R=R_g$ versus $V=\sqrt{3}\sigma_g$.  We assume all cluster potentials are dominated by dark matter, i.e., $f_b\simless 0.15$.

\bibliography{ref}

\label{lastpage}
\end{document}